\documentclass[sigconf]{acmart}

\AtBeginDocument{%
  }

\setcopyright{none}
\copyrightyear{2025}
\acmYear{2025}
\acmDOI{XXXXXXX.XXXXXXX}

\acmConference[]{}{}{}
\acmISBN{}

\begin{document}

\title{Artificial Empathy: AI based Mental Health}

\author{Aditya Naik}
\email{adinaik@iu.edu}
\affiliation{%
  \institution{Indiana University}
  \city{Indianapolis}
  \state{Indiana}
  \country{United States}
}

\author{Jovi Thomas}
\email{jovithom@iu.edu}
\affiliation{%
  \institution{Indiana University}
  \city{Indianapolis}
  \state{Indiana}
  \country{United States}
}

\author{Teja Sree Mandava}
\email{tmandava@iu.edu}
\affiliation{%
  \institution{Indiana University}
  \city{Indianapolis}
  \state{Indiana}
  \country{United States}
}

\author{Himavanth Reddy Vemula}
\email{vemreddy@iu.edu}
\affiliation{%
  \institution{Indiana University}
  \city{Indianapolis}
  \state{Indiana}
  \country{United States}
}

\begin{abstract}
Many people suffer from mental health problems but not everyone deems it necessary to see a professional and many don't have the means to see a mental health care provider. AI chatbots have increasingly become a go to for individuals who have mental disorders or would just like to talk about what's going on in their lives, so that they may be counseled through it. This paper presents a study and outcomes of participants who have used the chatbots before and a scenario based LLM model testing. Our findings indicate that AI chatbots were utilized for needs like a ``Five minute therapist'' or someone to just talk to. People also liked the fact that there is no judgement with the chatbots. However, many were and are still concerned about the privacy of their most vulnerable feelings and the security of their information. The testing of the LLMs raised a few concerns as well. There were some chatbots that were consistently reassuring, used emojis and names for a personal touch, and were quick to suggest professional help. But the models lacked in bridging the support gap, could have inconsistent tone or shift to inappropriate tones (casual or romantic), and could lack crisis sensitivity, failing to recognize red flag language or escalate responses appropriately. Our results can be utilized by both the technology and mental health care industries to notate and utilize AI chatbots for the purpose of helping individuals get through tough emotional times.
\end{abstract}



\maketitle

\section{Introduction}
Mental health can be defined as the absence of mental disease or it can be defined as a state of being that also includes the biological, psychological or social factors which contribute to an individual's mental state and ability to function within the environment \cite{carter1959proceedings}. Other definitions extend beyond this to also include intellectual, emotional and spiritual development, positive self-perception, feelings of self-worth and physical health, and intrapersonal harmony \cite{manwell2015mental}. Worldwide, Mental disorders are a leading cause of disability with significant economic, social, human rights, and health impacts. Mental health issues affect approximately one billion individuals worldwide each year. In the US alone, one in five adults experience mental illness annually, with a notable rise in recent decades characterized by increased rates of suicidal behavior, substance misuse, and social isolation \cite{han2025unleashing}. The global leap in mental disorders warrants scalable and innovative solutions for delivering therapy. More specifically, depression and anxiety are globally the most prevalent mental health disorders, affecting an estimated 322 million and 264 million individuals, respectively. Despite the escalating mental health demands, a global shortage of mental health professionals persists, with an unsustainable gap between the demand and supply of service providers \cite{rehm2019global,jeste2020battling}.

Individuals addressing mental health challenges go towards a variety of approaches which tailor to their specific needs and circumstances. Usually, these include psychotherapy, medication, support groups, and sometimes lifestyle modifications. The psychotherapy approach focuses on changing problematic behaviors, feelings, and thoughts. This is achieved by discovering their unconscious meanings and motivations. ``Psychoanalytic therapies emphasize a collaborative relationship between therapist and patient, helping individuals explore their thoughts, behaviors, and emotions to gain deeper self-awareness'' \cite{yuan2025improving}. Also, meeting others that may have faced similar challenges can offer emotional support and related advice. This is also where people share their experiences, coping mechanisms, and encouragement, thereby fostering a sense of community and understanding. In certain extreme conditions regarding mental health, medications are believed to be effective. Tyrosine, antidepressants, anti-anxiety medications, mood stabilizers, and antipsychotics are generally prescribed to regulate the symptoms \cite{yuan2025improving}. But despite all this, people who face mental issues are still found lost when it comes to treatment most of the time. This may be due to the availability of mental healthcare providers, the costs that are associated with such treatment, or even their personal circumstances. For example, in the United States, therapy costs can range from \$65 to \$250 per session, making it inaccessible for many individuals without insurance \cite{zhu2024insurance}.

Regarding Psychotherapy, finding a good psychiatrist may not be as easy as it sounds. A treatment for such a personalized problem can't have a generic solution. Medications come with side effects and also with a vicious cycle of placebo effects causing dependency. Some cultures discourage open discussions about mental health, making it harder for individuals to seek help in support groups. They might not even find such groups in the first place in these regions.

Some of the alternatives that the people tend to go towards are mostly based on a very inclusive environment. Sometimes, a solo path such as self help. Platforms like Reddit (r/mentalhealth), 7 Cups, and Discord support groups provide anonymous discussions and advice. Many individuals utilize cognitive behavioral therapy workbooks to manage symptoms. Apps such as Headspace (used for meditation), Moodpath (mood tracking), and Woebot (AI therapy) offer accessible self help tools. AI based mental health models also include chatbots like ELIZA. These rule based bots only use predefined scripts to simulate conversations. More advanced machine learning based models leverage Natural Language Processing and sentiment analysis. This is seen in Woebot where it analyzes emotions and provides personalized responses. Recently, large language models have been used for more nuanced interactions but raise concerns about data privacy and ethical considerations \cite{ai_mental_health_wiki,guo2024large}. Using scenario-based quick testing, we qualitatively examined the therapeutic efficacy and user experiences of twelve LLM-based mental health chatbots in order to investigate these issues.

These AI driven mental health tools have emerged as accessible alternatives to traditional therapy. They offer users on demand support through chatbots, mood tracking apps, and AI coaching systems \cite{bano2023effectiveness}. These technologies leverage cognitive behavioral therapy techniques, natural language processing, and sentiment analysis to simulate empathetic interactions \cite{molli2022effectiveness}. Experts in the mental health domain still feel effectiveness in truly understanding human emotions remains questionable \cite{ai_therapy_zenora}. Many AI models struggle with nuanced emotional recognition. This leads to false diagnoses leading to generic or misaligned responses that may not adequately support users in distress \cite{ai_therapy_abby}. Additionally, concerns about the data privacy of users, their security, and the potential biases embedded in AI algorithms raise ethical questions regarding their widespread adoption \cite{saeidnia2024ethical}. The emotional and therapeutic responsiveness of LLM-based AI mental health solutions is assessed in this study, which also identifies important drawbacks and design implications for future systems that are more inclusive and efficient.

\section{Related Work}

\subsection{State of the Cause}
With there being a stated global increase in mental disorders, it has become a probable necessity for the implementation of scalable and restructured mental therapy solutions. Large Language Model based mental health chatbots have swiftly emerged as an option for prevailing over the challenges of the cost, time and accessibility constraints that are often correlated with traditional mental health therapy \cite{guo2024large}. Recently, the rise of AI mediated care has piqued the interest of researchers comparing the traditional care method, the human, versus AI. While some of the research shows that AI chatbots are able to produce more empathetic and high quality responses than physicians \cite{perez2025ai,mahindru2023role}, other studies have found that users prefer empathetic responses, but by a human counterpart as opposed to AI \cite{kazdin2000encyclopedia}.

\subsection{Rise of LLM Based Mental Health Applications}
AI has raised both challenges and opportunities to better support mental healthcare interventions, with an undeniable growth of commercial AI--based mobile apps for mental health \cite{alotaibi2024review}. The use of LLMs in mental health is a relatively new and arising field. It has been found that there are few studies focused on LLM based applications. Most of the exploration of reviews have concentrated on the effectiveness of non LLM based mental health chatbots \cite{alotaibi2024review,yuan2025improving,abdAlrazaq2019overview,abdAlrazaq2020effectiveness,vaidyam2019chatbots}.

\subsection{Gaps in Current LLM Based Mental Health Research}
In addition, it had been found that the most pertinent LLM based mental health studies that have been identified thus far, are by \cite{guo2024large,hua2024large}, in which LLMs for mental health were reviewed but did not focus on LLM based chatbots. However, a more recent study, completed in January of 2024, reviews different LLM based models. In fact, the review identified 22 LLM based chatbots. They came to the conclusion that out of the 22 LLM based chatbots, the mental health chatbots could be categorized into five main groups: addressing general mental health, depression, anxiety, suicide ideation, and stress. In this study, they conclude that though the LLM based chatbots have great potential for empathetic interactions and therapeutic conversations; these tools still require continuous innovation. The critical gap in research that they identified was the lack of chatbots tailored to the unique needs of businesses and organizations, and perhaps students too \cite{guo2024large}. In contrast to previous evaluations, our study tests twelve LLM-based chatbots for mental health using prompts, evaluating their therapeutic potential through scenarios of real-world interactions.

\subsection{AI as Social and Emotional Companions}
As AI increasingly enters the communications process of the mental health space, taking on more social roles, it might be impossible to get around the fact that it is necessary and possibly inevitable for them to display mental and social capacities \cite{shao2023empathetic,breazeal2004designing,fong2003survey}. The development of social technologies challenges the assumption that social needs and seeking social support and companionship can only be fulfilled by a fellow human being. It also demonstrates that companion robots are capable of communicating emotions and are effective in providing emotional support, companionship, and psychological benefits \cite{shao2023empathetic,fitzpatrick2017delivering,robinson2013psychosocial,sabanovic2013paro,ta2020user}. Our study examines whether LLMs may deliver significant therapeutic responses that are in line with emotional depth and user trust, as measured against conventional treatment expectations, in contrast to earlier research that concentrated on AI's capacity to replicate companionship.

Though these studies have been conducted, the gap of data that we propose to fill is if the AI tools or more specifically chatbots, are not only able to provide emotional support and empathetic resolve but if they are able to give accurate mental health counsel as well. We also aim to fill in the data gap of AI based versus traditional mental healthcare by reviewing how AI based mental health models compare with traditional mental health therapy approaches.

\section{Problem Statement}
While traditional therapy provides personalized, human driven care, it comes with limitations such as accessibility and affordability \cite{molli2022effectiveness}. AI based tools, on the other hand, offer scalability but risk oversimplifying complex mental health needs \cite{saeidnia2024ethical}.

Given these considerations, it becomes crucial to explore the real impact of AI driven interventions. \textbf{Can they provide meaningful guidance which is comparable to human therapists? What are the challenges that the users face while using such an LLM model and what opportunities that can be implemented in these models.} Having thorough research on these topics will give experts that work in the field of Cognitive Large Language models and Mental health professionals an additional overview on the overall Artificial intelligence based mental health therapy.

\section{Methodology}

\subsection{Overview}
This study investigated whether large language model chatbots can provide morally sound and emotionally meaningful mental health assistance using a qualitative mixed methods approach. In order to learn about their experiences with chatbots, we first recruited participants using AI focused online platforms, then conducted questionnaires and interviews. Four emotionally complex situations were created using the insights from these replies as well as professional advice from a registered mental health practitioner. Eight LLM chatbots were assessed using these in real time dialogues. With expert validation to guarantee clinical relevance, we used affinity mapping and qualitative sentiment analysis to examine the bots' empathy, tone, and therapeutic efficacy.

\subsection{Participant Recruitment}
To assess whether AI chatbots are capable of providing emotionally relevant, encouraging, and morally good mental health advice, this study uses a qualitative mixed methods approach. Our approach was developed in many overlapping stages, starting with participant recruitment via targeted questionnaires distributed via AI focused Discord channels and Reddit mental health forums. The study used both closed ended and open ended questions designed to explore participants' familiarity with AI, their comfort level with discussing emotional topics with chatbots, and their willingness to engage in follow up interviews. We asked at the end of the survey if they are willing to have an interview and two participants came forward to give insights from their experience. We were able to determine user preferences and areas of concern about the use of LLMs in mental health situations.

\subsection{Survey and Interviews}
Following survey distribution, we conducted semi structured interviews with two student participants and one licensed mental health professional. We interviewed the students who use AI models to find help for their mental health. This improved our comprehension of how people behave, what they anticipate, and what their emotional requirements are when dealing with AI systems. When creating prompts for assessing mental health tools, the expert interview offered insights on therapeutic communication strategies, clinical practices, and possible warning signs. These discussions guided the development of contextually appropriate user prompts and helped us base our testing procedure on practical therapeutic reasoning.

\subsection{Testing and Evaluation}
Drawing on patterns identified in survey data, consultations with mental health professionals, and a review of commonly documented cognitive distortions in mental health literature, we developed four representative scenarios with prompts. These scenarios were designed to simulate realistic client concerns typically presented in therapeutic settings, making them well-suited for evaluating the responses of large language models. With this we were able to evaluate the bots' responsiveness, empathy, and therapeutic efficacy across a range of requirements by simulating real life discussions that people with different mental states may have with an AI. Each scenario reflected a distinct psychological profile. It began with a prompt introducing a fictional individual and their current emotional or situational context, thereby setting the stage for the AI model's response. These scenarios were created following an examination of typical cognitive distortions and mental health discourse patterns that are frequently seen in professional settings, such as underreporting emotional discomfort, disguising distress, and catastrophizing prompt was purposefully constructed to get more difficult emotionally as the exchange went on.

We tested the AI bot's ability to provide consoling replies and emotional presence by designing the ``Sad Story'' scenario around users who are obviously sad and expressly looking for validation and optimism. The fictional individual in the ``Sad Scenario'' is distressed and is self aware of their emotional condition. In contrast, the ``Normal Story'' introduces a user with a seemingly stable life who nonetheless experiences anxiety and self doubt, thereby testing whether the AI bot can detect subtle internal struggles beneath surface stability. In order to determine if the AI bot could recognize inconsistencies and provide more profound psychological insights, the ``Mixed Story'' presented a more complicated task. It included a user who thought they were emotionally stable but revealed fragmented and unhealthy thought habits. This scenario was intended to assess the models' ability to challenge the users maladaptive beliefs and redirect them towards a more constructive and healthier decision making pathways. Lastly, the ``Moderate Story'' highlighted a user who, despite their positive demeanor, conceals emotional instability and mood swings. The AI bot's capacity to detect concealed sadness and react with sophisticated empathy was put to the test in this instance. When taken as a whole, these thoughtfully crafted questions enabled us to examine not just the AI's speech but also its perceptions and reactions to various emotional truth layers, a crucial aspect of assessing the reliability and security of AI powered mental health resources.

Twelve LLM based chatbots were selected based on the survey. The selection consisted of a mix of mainstream AI chatbots along with AI chatbots that are specifically designed for Mental health. ChatGPT, Gemini, Replika, Anima, Character AI (Therapist personality), Woebot, Antar, Wysa, Claude, Perplexity, Grok and Deepseek were used to test these prompts. In order to maintain the flow of a normal discourse about mental health, the interactions were carried out in real time, with each prompt being submitted in turn. During each session, prompt observations were made and responses were entered into a standardized spreadsheet. We used affinity mapping to group emotional themes, reaction patterns, and recurrent success or failure patterns in the bots' outputs after the tests were over. To evaluate how each AI system interacted with the user's emotional state and their capacity to have morally sound and encouraging conversations, we also carried out a qualitative sentiment and tone study.

\subsection{Follow up with mental health professional}
Lastly, the mental health specialist examined a few transcripts and provided comments about the suitability of the bots' crisis sensitivity, validation methods, and therapeutic language. This outside validation aided in placing the findings into accepted frameworks for mental health and helped us formulate design suggestions for enhancing trauma sensitive AI systems. The emotional demands of users, the functional constraints of existing LLMs, and the possible future roles these systems may play in enhancing conventional treatment were all connected through this approach.

\section{Results}
The data gathered from participant surveys, expert interviews, and iterative AI model testing offered multiple insights for our central research question. Our affinity mapping process distilled the findings into five key thematic categories. Each category highlights distinct dimensions of the user experiences, model efficacy, and areas for future enhancement for AI driven mental health support.

\subsection{Participant Preferences}

\begin{figure}[h]
  \centering
  \includegraphics[width=\linewidth]{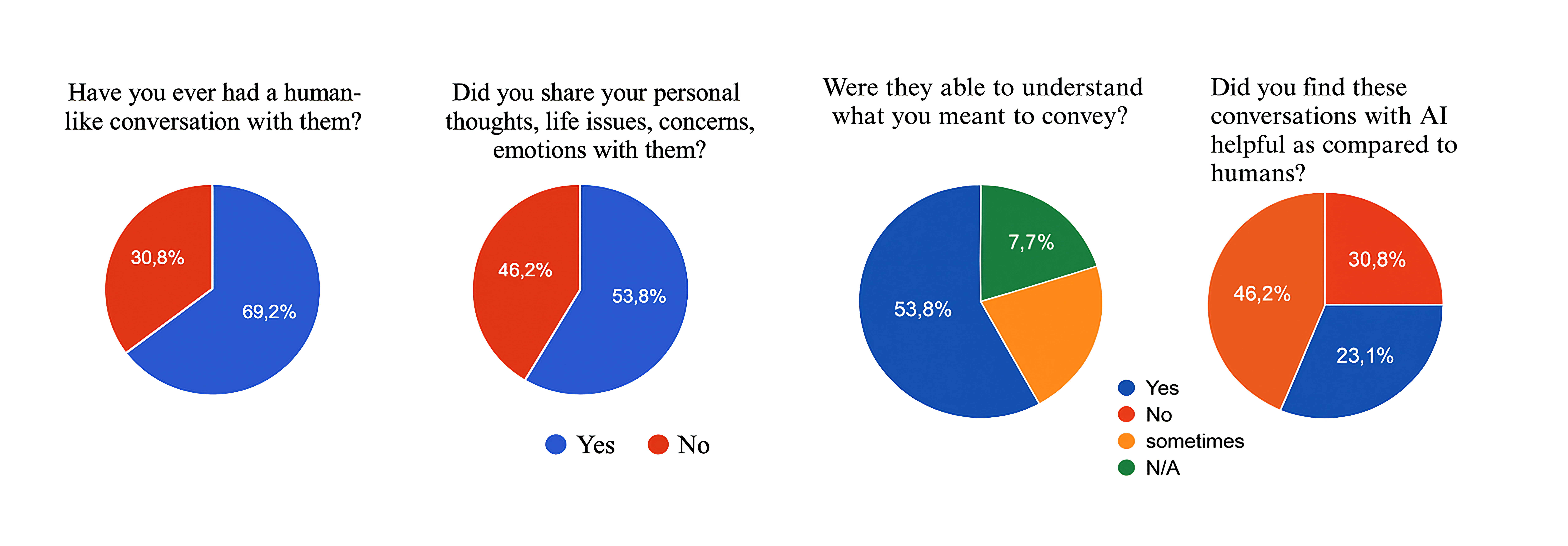}
  \caption{Pie charts showing participants' insights on LLM usage and preferences.}
  \Description{Four pie charts showing survey results about participants' human-like conversations with AI, sharing personal information, AI's understanding capability, and helpfulness compared to humans.}
  \label{fig:participant_insights}
\end{figure}

Our findings are based on a study involving 15 participants, including one from the mental health profession. The study primarily attracted younger adult participants aged between 20 to 30 years. This suggested that younger individuals may be more receptive to new technology and modern treatment solutions such as the AI based therapy. This inclination may also reflect underlying social dynamics. This could be the absence of supportive interpersonal relationships or trusted confidants among younger individuals. Interestingly, despite including specialized mental health oriented AI models such as Anima and Wysa, participants showed a clear preference for mainstream LLMs like ChatGPT and Deepseek. This data points to a potential correlation between model accessibility, familiarity, and perceived performance.

\subsubsection{Patterns of LLM Usage}
Participants frequently used LLMs for tasks across three broad categories. First was Writing (e.g. grammar corrections, composing emails), Academic and Research Support (e.g. analyzing data, generating research content), and Creative and Exploratory Tasks (e.g. brainstorming ideas, normal conversating). Notably, these tasks were characterized by their non-judgmental nature. While not initially designed for therapeutic purposes, the supportive, informational, and non judgmental nature of these interactions underscores the potential of general LLMs in complementary mental health roles.

\subsubsection{User Experiences and Emotional Sharing}
A significant proportion of participants (69.2\%) reported engaging in conversational interactions with LLM models as if they were human. Approximately 53\% disclosed personal thoughts, emotions, and life concerns. However, privacy and emotional depth concerns discouraged many from deeper engagement, which highlighted a critical barrier. The participants expressed apprehensions about data security and perceived a lack of genuine emotional empathy in the AI responses.

\subsubsection{Perceived Empathy and Effectiveness}
Participants exhibited mixed feelings regarding their emotional connections with AI. While some appreciated the immediacy, non judgmental interaction, and convenience some even described the AI as a ``five minute therapist,'' others felt that AI lacked emotional depth. They pointed towards the responses being monotonous, or generic and repetitive. Approximately half of the participants indicated uncertainty about the effectiveness of AI compared to a real human.

\subsection{Empathic Responses}
Systematic testing across multiple AI platforms highlighted several insights. Table~\ref{tab:llm_comparison} summarizes core aspects of mental health support tested with the LLM models. These attributes include empathy, validation, crisis guidance, and tone offered by each LLM model. We were unable to test our prompts with Anima, Character AI, Claude, and Grok due to access restrictions and paywalls. This represents a limitation in our study, which we discuss in the following sections.

\begin{table*}
  \caption{Mapping of LLM Model Testing Data}
  \label{tab:llm_comparison}
  \small
  \begin{tabular}{p{1.3cm}p{1.6cm}p{1.8cm}p{1.5cm}p{1.7cm}p{1.5cm}p{1.3cm}p{1.6cm}p{1.5cm}}
    \toprule
    \textbf{LLM} & \textbf{Opening Tone} & \textbf{Empathy Strategy} & \textbf{Probing Questions} & \textbf{Validation Style} & \textbf{Crisis Navigation} & \textbf{Tone Markers} & \textbf{Coping Suggestions} & \textbf{Patterns}\\
    \midrule
    ChatGPT & Warm, validating & Emotional reassurance & Yes; reflective & Strong validation & Suggests external help & Emojis; conversational & Self-care, talk to someone & Highly structured\\
    \midrule
    Gemini & Compassionate, balanced & Normalizes feelings & Yes; explores roots & Empathetic, realistic & Recommends professional help & Minimal; professional & Maintain structure, talk to someone & Clinical + friendly\\
    \midrule
    Replika & Casual, friendly & Self-references, encouragement & Minimal; light & Mild reassurance & Rare; avoids depth & Nickname use; quirky & Limited & Companionship focus\\
    \midrule
    DeepSeek & Highly empathetic & Compassion + psychological framing & Yes; explores perceptions & Strong; challenges distortions & Strongly encourages outreach & Emoji; metaphors & Naming, reframing thoughts & Therapeutic + conversational\\
    \midrule
    Perplexity & Gently supportive & Presence, shared space & Yes; invites exploration & Reassures, validates & Encourages outreach & Conversational; warm & Express feelings & Gentle listener\\
    \bottomrule
  \end{tabular}
\end{table*}

Models such as Claude and ChatGPT demonstrated consistent emotional validation. They included therapeutic communication techniques such as affect labeling, gentle reframing, and trauma informed language. DeepSeek and ChatGPT further personalized interactions through the use of names, emojis. Claude, in particular, excelled in mirroring the tone and talking style commonly used by human counselors. On the structural side, Antar, and Wysa provided responses which are rooted in cognitive behavioral therapy (CBT). This method offers goal oriented support frameworks and low barrier environments for emotional expression. Perplexity stood out for its provision of psychoeducational content and the inclusion of citations. While it enhanced the perceived credibility of its suggestions, the addition of this felt pointless in terms of a normal human conversation.

\subsection{Challenges and Limitations}
While the responses provided by the AI models were generally credible, they exhibited recurring limitations that constrained their overall effectiveness. It started with accessibility of the AI models. The models, Anima, Wysa had multiple steps in order to start with the conversation. In the case of Woebot, the accessibility was very limited which led to no testing at all. For Anima and Grok, hiccups were faced during prolonged sessions as the models had paywall and required the user to pay to continue talking. Popular and mainstream AI models, such as ChatGPT and DeepSeek, offered significantly easier access, often not requiring login credentials or account setup. From our surveys, we have realized that access plays a crucial factor in mental health therapy and these patterns underscore a concerning accessibility gap. Mental health support becomes more effective only when users can access and afford advanced AI features. This also raises important design questions around equity and inclusion. Future mental health tools must ensure that meaningful emotional support is not locked behind premium paywalls, which may exclude vulnerable users in need of consistent, high quality care. The current tiered service model risks creating emotional inequities, where those unable to pay face reduced or superficial support compared to paying users.

Wysa, on the other hand, relied on preset multiple choice questions for nearly any prompt the user entered. This significantly limited users' freedom to express themselves, as it removed the option to type out their responses. Multiple models (Character AI, Perplexity) failed to consistently recognize emotional cues or cognitive distortions. This made them spiral down the wrong path, offering responses that lacked empathy or therapeutic nuance. Character AI for instance would focus on one part of the prompt and continue to expand that concern even when it wasn't the overall problem with the user. Several models also exhibited inadequate crisis sensitivity. They failed to identify critical emotional red flags and provided inappropriate, overly casual, or trivializing responses during serious emotional disclosures. Models such as Claude and ChatGPT demonstrated relatively stronger emotional validation and reflective communication. Structured cognitive behavioral therapy based models like Antar, and Wysa provided more methodically structured support but still faced limitations in emotional depth and crisis response. There were issues noted with the consistency and continuity of the responses. The participants noted frequent inconsistencies in tone and a lack of contextual memory across sessions. This caused reduced perceived emotional investment and relational continuity. These findings suggest that although current LLMs can replicate certain elements of empathic support and structured therapy, they fall short of delivering holistic, context sensitive guidance akin to that provided by trained human professionals.

\subsection{Opportunities for Improvement}
Participant feedback and professional evaluations identified clear avenues for enhancing AI driven mental health tools. Participants stressed the importance of accessible platforms that do not unnecessarily request personal information and location, thus safeguarding user privacy. Implementing memory and relational continuity within AI models could substantially improve user experience, allowing follow up interactions and deeper emotional tracking. Integrating nuanced language, trauma informed strategies, and crisis de-escalation capabilities emerged as crucial needs to better handle sensitive emotional contexts and ensure appropriate escalation pathways to human support when necessary.

\section{Discussion}
The results of this study confirm our initial hypothesis: individuals often avoid traditional mental health services due to barriers such as cost, limited access, and social stigma. In such circumstances, many turn to LLM based tools as a form of self guided mental health support.

\begin{table*}
  \caption{Characteristics of AI Based Mental Health Tools}
  \label{tab:characteristics}
  \small
  \begin{tabular}{p{3.2cm}p{3.2cm}p{3.2cm}p{3.2cm}}
    \toprule
    \textbf{Characteristic} & \textbf{Strength} & \textbf{Weakness} & \textbf{Improvement}\\
    \midrule
    Emotional Validation & Consistent validation & Generic reassurances, casual tones & Nuanced, trauma-informed language\\
    \midrule
    Therapeutic Approach & CBT-based frameworks & Inadequate crisis sensitivity & Crisis de-escalation capabilities\\
    \midrule
    Contextual Understanding & Psychoeducational content & Inconsistencies, lack of memory & Memory and continuity\\
    \midrule
    Privacy & Low-barrier expression & Failure to bridge to human help & Accessible platforms, less personal data\\
    \bottomrule
  \end{tabular}
\end{table*}

But according to our research, these tools usually fail to provide contextually rich or emotionally responsive responses, which makes it hard for users to feel understood or supported. This conclusion aligns with the body of literature that emphasizes how, despite their conversational fluency, contemporary AI systems frequently lack the emotional complexity and depth required to support mental health \cite{ai_therapy_zenora,saeidnia2024ethical}. Beyond confirmation of the hypothesis, several novel behavioral insights emerged. Users often treat AI as a form of ``five minute therapist,'' using it to process immediate emotions or seek quick advice, a behavior noted in related studies examining casual use of AI for emotional support \cite{iftikhar2024reimagining,ayers2023comparing}. Additionally, participants reported using AI to confirm decisions, engage in internal debates, or simulate social interactions by assigning personas to the AI system. These practices reveal that, while not always reliable for deep emotional engagement, AI is filling a functional gap in users' self-care routines \cite{molli2022effectiveness}.

Several design implications arise from our study. First, accessibility remains paramount AI tools should not require sensitive personal data unless necessary for diagnosis \cite{ai_therapy_abby}. Second, implementing memory or contextual tracking can simulate relational continuity, enhancing emotional trust, an aspect shown to increase user engagement in conversational agents \cite{fong2003survey,ta2020user}. Third, trauma informed design practices are essential to avoid harm, especially among at-risk populations \cite{saeidnia2024ethical,robinson2013psychosocial}. Furthermore, escalation protocols to human therapists or hotlines must be integrated, particularly during moments of crisis \cite{abdAlrazaq2019overview}. While some systems (e.g. Perplexity) demonstrate partial implementation of such features, broader adoption remains limited. Lastly, the therapeutic flexibility of AI needs to be improved. Tools should offer a variety of treatment approaches catered to different user backgrounds rather than merely relying on Cognitive Behavioral Therapy's (CBT) recommendations \cite{fitzpatrick2017delivering}. Additionally, tone modulation is critical. AI must distinguish between casual and serious interactions to avoid emotional misalignment. Systems should be capable of adjusting tone and responses based on the user's progress over time \cite{ayers2023comparing}. In sum, while AI systems offer convenience and immediacy, they currently lack the emotional intelligence required for effective mental health support. Addressing these limitations will be essential for the development of safer, more empathetic, and user aligned mental health technologies.

Additionally, during a second consultation with the mental health professional but this time informed by our testing findings, it was noted that while the responses generated by the models were often impressive, they still lacked the nuanced sensitivity which is typically offered by human therapists. The professional emphasized that LLMs could serve as supportive companions rather than direct replacements for mental health practitioners \cite{kazdin2000encyclopedia}. They pointed out a critical limitation in AI's personalization: ``AI only gets better with its responses the more it is used by the client.'' The fact that the efficacy of these systems is largely dependent on frequent usage and the collection of user specific data highlights a fundamental problem in therapeutic AI. AI is unable to understand tiny indicators like body language and changes in speech tone holistically, unlike human therapists who can do so instantly \cite{morris2018towards}. LLMs must thus be constructed with feedback loops and escalation processes to guarantee that users in distress are transferred to human assistance when necessary, even though they may be useful for early engagement or continuing support.

\section{Conclusion}
In this paper, we have presented research of LLM models and their therapeutic effects on mental health. From this research we learned that AI chatbots can offer valuable, immediate support to individuals, especially those seeking non-judgmental and easily accessible interactions. However, there are still concerns about privacy, the level of emotional understanding of the chatbots, and the lack of genuine and human empathy. Many users appreciate the convenient nature and non-judgmental association with AI support, while others remain concerned that it may feel impersonal during times of real emotional upset and vulnerability. As mental health cases rise, it is becoming increasingly hard for mental health professionals to keep up with the demand. AI chatbots are a necessary resource to supplement mental health care. According to our research, AI chatbots need to be developed in many areas if they are to be genuinely useful. These include the capacity to maintain relational continuity throughout sessions, use communication techniques inspired by trauma, and provide insightful, situation-specific advice. We also underscore how crucial it is to put crisis escalation procedures and feedback loops in place so that users who are experiencing difficulties may be easily sent to human experts when necessary. Although LLMs have the potential to promote mental well-being, this research adds to the design and ethical discussion around mental health technologies by demonstrating that they need to be created with increased emotional intelligence, customization, and accountability.

\section{Future Work}
There is still much to be done in the study and effort of recognizing and concluding if AI chatbots can truly deliver meaningful mental health support. Perhaps a more comprehensive study, which would involve a larger and more diverse participant pool, with an extended timeframe will be necessary and ideal to be successful with this research. Because individuals' responses to AI mental health interventions vary greatly, some appreciate the accessibility and impartiality of chatbots while others remain distrustful of the bots' ability to legitimately understand the emotions and experiences of a human being, further research is a must to address these nuances and to better assess and understand AI's role in mental healthcare. Another method of research concerns an individual's level of accessibility. Some may not have the ability to pay for the premium version of some LLM models so, would having the ability to pay for premium services render better, more adaptive and dynamic responses? Also, the implementation and evolution of relational continuity is imperative and deserves exploration and investigation in the near future. Further, more in depth research is due to be able to answer such questions and concerns. Ultimately, answering these questions is essential to ensure equitable, emotionally safe AI mental health tools for all user populations.

\bibliographystyle{ACM-Reference-Format}
\bibliography{mental-health-refs}

\begin{thebibliography}{10}

\bibitem{abdAlrazaq2019overview}
{\sc Abd-Alrazaq, A.~A., Alajlani, M., Alalwan, A.~A., Bewick, B.~M., Gardner,
  P., and Househ, M.}
\newblock An overview of the features of chatbots in mental health: A scoping
  review.
\newblock {\em International Journal of Medical Informatics\/} (2019).

\bibitem{abdAlrazaq2020effectiveness}
{\sc Abd-Alrazaq, A.~A., Rababeh, A., Alajlani, M., Bewick, B.~M., and Househ,
  M.}
\newblock Effectiveness and safety of using chatbots to improve mental health:
  Systematic review and meta-analysis.
\newblock {\em Journal of Medical Internet Research 22}, 7 (2020), e16021.

\bibitem{alotaibi2024review}
{\sc Alotaibi, A., and Sas, C.}
\newblock Review of ai-based mental health apps.
\newblock In {\em Proceedings of the 36th International BCS Human-Computer
  Interaction Conference (BCS HCI '23)\/} (Swindon, GBR, 2024), BCS Learning
  \& Development Ltd, pp.~238--250.

\bibitem{ayers2023comparing}
{\sc Ayers, J.~W., Poliak, A., Dredze, M., Leas, E.~C., Zhu, Z., Kelley,
  J.~B., Faix, D.~J., Goodman, A.~M., Longhurst, C.~A., Hogarth, M., and
  others}.
\newblock Comparing physician and artificial intelligence chatbot responses to
  patient questions posted to a public social media forum.
\newblock {\em JAMA Internal Medicine\/} (2023).

\bibitem{bano2023effectiveness}
{\sc Bano, H.}
\newblock The effectiveness of ai-based therapy -- is it better than
  traditional psychotherapy?
\newblock 2023.

\bibitem{breazeal2004designing}
{\sc Breazeal, C.}
\newblock {\em Designing sociable robots}.
\newblock MIT Press, 2004.

\bibitem{carter1959proceedings}
{\sc Carter, L.~F.}
\newblock Proceedings of the sixty-seventh annual business meeting of the
  american psychological association, inc.: Report of the recording secretary.
\newblock {\em American Psychologist 14}, 12 (1959), 741--763.

\bibitem{fitzpatrick2017delivering}
{\sc Fitzpatrick, K.~K., Darcy, A., and Vierhile, M.}
\newblock Delivering cognitive behavior therapy to young adults with symptoms
  of depression and anxiety using a fully automated conversational agent
  (woebot): a randomized controlled trial.
\newblock {\em JMIR Mental Health 4}, 2 (2017), e7785.

\bibitem{fong2003survey}
{\sc Fong, T., Nourbakhsh, I., and Dautenhahn, K.}
\newblock A survey of socially interactive robots.
\newblock {\em Robotics and Autonomous Systems 42}, 3-4 (2003), 143--166.

\bibitem{guo2024large}
{\sc Guo, Z., Lai, A., Thygesen, J.~H., Farrington, J., Keen, T., and Li, K.}
\newblock Large language model for mental health: A systematic review.
\newblock {\em arXiv preprint arXiv:2403.15401\/} (2024).

\bibitem{han2025unleashing}
{\sc Han, Q., and Zhao, C.}
\newblock Unleashing the potential of chatbots in mental health: Bibliometric
  analysis.
\newblock {\em Frontiers in Psychiatry\/} (2025).

\bibitem{hua2024large}
{\sc Hua, Y., Liu, F., Yang, K., Li, Z., Sheu, Y.-H., Zhou, P., Moran, L.~V.,
  Ananiadou, S., and Beam, A.}
\newblock Large language models in mental health care: A scoping review.
\newblock {\em arXiv preprint arXiv:2401.02984\/} (2024).

\bibitem{iftikhar2024reimagining}
{\sc Iftikhar, Z.}
\newblock Re-imagining mental health access: The role of human, ai and design.
\newblock In {\em Companion Publication of the 2024 Conference on
  Computer-Supported Cooperative Work and Social Computing (CSCW Companion
  '24)\/} (New York, NY, USA, 2024), Association for Computing Machinery,
  pp.~57--60.

\bibitem{jeste2020battling}
{\sc Jeste, D.~V., Lee, E.~E., and Cacioppo, S.}
\newblock Battling the modern behavioral epidemic of loneliness: Suggestions
  for research and interventions.
\newblock {\em JAMA Psychiatry 77}, 6 (2020), 553--554.

\bibitem{kazdin2000encyclopedia}
{\sc Kazdin, A.~E.}, Ed.
\newblock {\em Encyclopedia of Psychology}.
\newblock American Psychological Association, 2000.

\bibitem{mahindru2023role}
{\sc Mahindru, A., Patil, P., and Agrawal, V.}
\newblock Role of physical activity on mental health and well-being: A review.
\newblock {\em Cureus 15}, 1 (2023).

\bibitem{manwell2015mental}
{\sc Manwell, L.~A., Barbic, S.~P., Roberts, K., Durisko, Z., Lee, C., Ware,
  E., and McKenzie, K.}
\newblock What is mental health? evidence towards a new definition from a mixed
  methods multidisciplinary international survey.
\newblock {\em BMJ Open 5}, 6 (2015), e007079.

\bibitem{molli2022effectiveness}
{\sc Molli, V. L.~P.}
\newblock Effectiveness of ai-based chatbots in mental health support: A
  systematic review.
\newblock {\em Journal of Healthcare AI and ML 9}, 9 (2022), 1--11.

\bibitem{morris2018towards}
{\sc Morris, R.~R., Kouddous, K., Kshirsagar, R., and Schueller, S.~M.}
\newblock Towards an artificially empathic conversational agent for mental
  health applications: system design and user perceptions.
\newblock {\em Journal of Medical Internet Research 20}, 6 (2018), e10148.

\bibitem{perez2025ai}
{\sc Perez, A.~L.}
\newblock Ai in therapy: Exploring its benefits and limitations for mental
  health.
\newblock Carepatron.com, 2025.

\bibitem{rehm2019global}
{\sc Rehm, J., and Shield, K.~D.}
\newblock Global burden of disease and the impact of mental and addictive
  disorders.
\newblock {\em Current Psychiatry Reports 21}, 2 (2019), 10.

\bibitem{robinson2013psychosocial}
{\sc Robinson, H., MacDonald, B., Kerse, N., and Broadbent, E.}
\newblock The psychosocial effects of a companion robot: a randomized
  controlled trial.
\newblock {\em Journal of the American Medical Directors Association 14}, 9
  (2013), 661--667.

\bibitem{sabanovic2013paro}
{\sc {\v{S}}abanovi{\'c}, S., Bennett, C.~C., Chang, W.-L., and Huber, L.}
\newblock Paro robot affects diverse interaction modalities in group sensory
  therapy for older adults with dementia.
\newblock In {\em 2013 IEEE 13th international conference on rehabilitation
  robotics (ICORR)\/} (2013), IEEE, pp.~1--6.

\bibitem{saeidnia2024ethical}
{\sc Saeidnia, H.~R., Hashemi~Fotami, S.~G., Lund, B., and Ghiasi, N.}
\newblock Ethical considerations in artificial intelligence interventions for
  mental health and well-being: Ensuring responsible implementation and impact.
\newblock {\em Social Sciences 13}, 7 (2024), 381.

\bibitem{shao2023empathetic}
{\sc Shao, R.}
\newblock An empathetic ai for mental health intervention: Conceptualizing and
  examining artificial empathy.
\newblock In {\em Proceedings of the 2nd Empathy-Centric Design Workshop
  (EmpathiCH '23)\/} (New York, NY, USA, 2023), Association for Computing
  Machinery, pp.~1--6.

\bibitem{ta2020user}
{\sc Ta, V., Griffith, C., Boatfield, C., Wang, X., Civitello, M., Bader, H.,
  DeCero, E., Loggarakis, A., and others}.
\newblock User experiences of social support from companion chatbots in
  everyday contexts: thematic analysis.
\newblock {\em Journal of Medical Internet Research 22}, 3 (2020), e16235.

\bibitem{vaidyam2019chatbots}
{\sc Vaidyam, A.~N., Wisniewski, H., Halamka, J.~D., Kashavan, M.~S., and
  Torous, J.~B.}
\newblock Chatbots and conversational agents in mental health: A review of the
  psychiatric landscape.
\newblock {\em Canadian Journal of Psychiatry 64}, 7 (2019), 456--464.

\bibitem{ai_mental_health_wiki}
{\sc Wikipedia}.
\newblock Artificial intelligence in mental health, 2023.

\bibitem{yuan2025improving}
{\sc Yuan, A., Colato, E.~G., Pescosolido, B., Song, H., and Samtani, S.}
\newblock Improving workplace well-being in modern organizations: A review of
  large language model-based mental health chatbots.
\newblock {\em ACM Transactions on Management Information Systems 16}, 1
  (2025), 3.

\bibitem{ai_therapy_abby}
{\sc {Abby.gg}}.
\newblock Ai therapy: Exploring its effectiveness and real-world applications,
  2025.

\bibitem{ai_therapy_zenora}
{\sc {Zenora.app}}.
\newblock Ai vs. traditional therapy: Comparing outcomes and experiences, 2024.

\bibitem{zhu2024insurance}
{\sc Zhu, J.~M., Huntington, A., Haeder, S., Wolk, C., and McConnell, K.~J.}
\newblock Insurance acceptance and cash pay rates for psychotherapy in the us.
\newblock {\em Health Affairs Scholar 2}, 9 (2024), qxae110.

\end{thebibliography}

\end{document}